\documentclass[prd,aps,epsf,amsfonts,floats,twocolumn,amssymb,amsmath,groupedaddress,showpacs,floatfix,nofootinbib]{revtex4}

\usepackage[]{latexsym}
\usepackage{bm}
\usepackage{subeqnarray}

\usepackage{graphicx,amssymb,amsmath}
\usepackage[english]{babel}
\usepackage{graphics}                 
\usepackage{color}                    
\usepackage{hyperref}

\newcommand{\be}{\begin{equation}}\newcommand{\ee}{\end{equation}}
\newcommand{\bea}{\begin{eqnarray}}\newcommand{\eea}{\end{eqnarray}}
\newcommand{\bsa}{\begin{subeqnarray}}
\newcommand{\esa}{\end{subeqnarray}}
\newcommand{\brr}{\begin{array}}\newcommand{\err}{\end{array}}
\newcommand{\bit}{\begin{itemize}}\newcommand{\eit}{\end{itemize}}
\newcommand{\ben}{\begin{enumerate}}\newcommand{\een}{\end{enumerate}}

\newcommand{\ba}{\begin{array}}
\newcommand{\ea}{\end{array}}

\def\lf{\left}

\def\non{\nonumber}

\def\ri{\right}

\def\1{{_{1}}}\def\2{{_{2}}}

\def\noHe0{:\;\!\!\;\!\!:H_e(0):\;\!\!\;\!\!:}
\def\noHm0{:\;\!\!\;\!\!:H_\mu(0):\;\!\!\;\!\!:}

\def\lf{\left}

\def\non{\nonumber}

\def\ri{\right}

\def\1{{_{1}}}\def\2{{_{2}}}

\begin{document}

\title{Vacuum condensate, geometric phase, Unruh effect and temperature measurement}

\author{ Antonio Capolupo${}^{\flat}$}
\author{ Giuseppe Vitiello${}^{\flat}$}
   \affiliation{${}^{\flat}$
  Dipartimento di Fisica E.R.Caianiello
  Universit\'a di Salerno, and INFN Gruppo collegato di Salerno, Fisciano (SA) - 84084, Italy}


\begin{abstract}

In our previous work it has been shown the possibility to use the Aharonov-Anandan invariant  as a tool in the analysis of disparate systems, including Hawking and Unruh effects, as well as graphene  physics and thermal states. We show that the vacuum condensation, characterizing such systems, is also related with geometric phases and
we analyze the properties of the geometric phase of systems represented by mixed state and undergoing a nonunitary evolution.  In particular, we consider two level atoms accelerated by an external potential and interacting with a thermal state. We propose the realization of Mach-Zehnder interferometers which can prove the existence of the Unruh effect and can allow very precise measurements of temperature.

\end{abstract}

\maketitle

\section{ Introduction}

It is hard to observe  phenomena like Unruh \cite{Unruh:1976db}, Hawking \cite{Hawking:1974sw} and Parker effects \cite{Parker:1968mv,Schrodinger}.
However it has been shown \cite{Capolupo:2013xza} that in such phenomena as well as in all the systems where the vacuum condensates are generated \cite{Takahashi:1974zn}-\cite{ Birrell}, the Aharonov--Anandan invariant (AAI)  \cite{Anandan:1990fq} is produced
 in their evolution.
Moreover, it has been shown \cite{Pati1,Pati2} that such an invariant  is related to the geometric phase \cite{Berry:1984jv}--\cite{Tong}. This fact suggests that in all the above phenomena,  in which the presence of AAIs has been revealed \cite{Capolupo:2013xza},  an associated  geometric phase also appears. Then, the study of such a phase could  open a   new way to the detection of effects elusive to the detection.

Here, instead of using AAI, which is experimentally difficult to be
observed, we use  geometric phases to study the Unruh effect and the possibility to perform very precise measurement of temperature.
Geometric phases have been detected in many physical systems \cite{Tomita}-\cite{Pechal} and have been also related to  CPT symmetry  \cite{Capolupo:2011rd} and
  SUSY violation in thermal states \cite{Capolupo:2013ica}.

 In the present paper, we focus our attention on systems of atoms   accelerated in an electromagnetic field and  atoms interacting with thermal states, so that we study   geometric phases for mixed states in
nonunitary, noncyclic evolution.
Different approaches exist in dealing with geometric phases and mixed states \cite{Uhlmann}-\cite{Tong}; in this paper we mostly follow the Wang and Liu approach \cite{Wang}.

In our treatment,
 besides the relation between boson condensation and geometric phase, 
we obtain two novel results, one concerning atoms accelerated in an electromagnetic field and the Unruh effect, the other one exhibiting the possibility to perform very precise measurement of temperature. In the first case, a detectable difference of the geometric phases appears between the accelerated and the inertial atoms. Such a phase difference is due only to the Unruh effect.
In the second case, the difference between geometric phases produced by atoms interacting with two different thermal states allows to determine the temperature of a sample once the temperature of the other one (assumed as reference temperature) is known.

The idea of using geometric phases and invariants to probe the Unruh effect  \cite{Capolupo:2013xza,Ivette1,Hu-Yu},
and to  build a thermometer \cite{Capolupo:2013xza,Ivette} has been already presented in  previous works,
(the study of dynamical phase to have precise estimation of the temperature has been proposed in \cite{IvetteF}).
In this paper we consider a realistic scheme for experimental implementations and we study the geometric phase defined in \cite{Wang},  which generalize the Berry phase (used in \cite{Ivette1,Ivette}), to the case of quantum open systems.
 The use of the phase presented in \cite{Wang} allows to consider time intervals arbitrary small (we do not need to consider  cyclic evolutions and their related period), and to take into account transition frequencies very low and spontaneous emission rates characteristic of fine and hyperfine structure of the atoms.
Indeed, in the short time intervals which we consider, the number of spontaneously emitted particle is negligible and the systems are quasi-stable.

The Berry phase for mixed states cannot be used  in the study of  the systems which we consider. One reason is that     our systems do not undergo cyclic evolutions because of their interaction with the environment \cite{Wang}. Even in the approximation of a quasicyclicity of mixed states, the number of spontaneously emitted particles becomes not  negligible  in time intervals of the order of $T = 2 \pi /\omega_{0}$ (with $\omega_{0}$ atomic transition frequency). Thus   the analysis of the geometric phase loses meaning.

Other improvements, out of reach  when using  Berry phase, offered by our treatment are that considering the fine and hyperfine structure of atoms allows to lower accelerations, which in turn improves the detection of Unruh effect and permits  very precise temperature measurements.

We consider  the structure of the atomic levels of $^{85}Rb$, $^{87}Rb$ and  $^{133}Cs$. We suggest that a Mach-Zehnder interferometer of $4 cm$, where the hyperfine level of these atoms is considered, may reveal  the  Unruh effect with accelerations of order of $10^{16}m/s^{2}$ and that a similar device could be used also for very precise temperature measurements.

In Section II we show the links relating AAI, geometric phase and vacuum condensates and introduce the geometric phase for mixed states in non-unitary evolution.  In Section III we analyze the geometric phase for a two level atom undergoing a non-unitary evolution and in Sections IV and V we study the realization of a Mach-Zehnder interferometer to reveal the Unruh effect and to build a very precise thermometer. Section VI is devoted to the conclusions.

\section{ Geometric phase }

The geometric phase for pure states can be expressed as \cite{Pati2}
 \bea \label{geom}\non
 \Phi(\Gamma) & = & \int_{\Gamma} \sqrt{ d D^{2} - d S^{2}}
\\
& = & \int_{\Gamma} \lf[  \lf \langle \frac{\phi}{\|\phi \|} \lf | i \frac{d}{dt} -  \dot{\psi}(t) \ri |  \frac{\phi}{\|\phi \|}\ri \rangle \ri] dt \,,
 \eea
which also establishes the relation with the AAI. Here $dD^2$ denotes the infinitesimal "reference distance" in the projective Hilbert space $P$ and is given by
\bea\non\label{refdist}
dD^{2} &=& \Big[\lf\langle \frac{d}{dt}\lf(\frac{\phi}{\|\phi \|} \ri)\Big |\frac{d}{dt}\lf(\frac{\phi}{\|\phi \|} \ri) \ri\rangle  + \dot{\psi}^{2}
\\
& - & 2i \dot{\psi}
\lf\langle  \frac{\phi}{\|\phi \|}  \Big |\frac{d}{dt}\lf(\frac{\phi}{\|\phi \|} \ri) \ri\rangle \Big] dt^{2}\,,
\eea
with
$
\psi(t)=\frac{i}{2}\ln \lf[\frac{\langle \phi (0)|\phi(t)\rangle}{\langle \phi (t)|\phi(0)\rangle} \ri].
$
In Eq.(\ref{geom}), $ d S^{2}$ is the Fubini-Study metric given by
 \bea\non\label{Fubdist}
dS^{2} &=& \Big(\lf\langle \frac{d}{dt}\lf(\frac{\phi}{\|\phi \|} \ri)\Big |\frac{d}{dt}\lf(\frac{\phi}{\|\phi \|} \ri) \ri\rangle
\\
& - & \lf [i
\lf\langle  \frac{\phi}{\|\phi \|}  \Big |\frac{d}{dt}\lf(\frac{\phi}{\|\phi \|} \ri) \ri\rangle \ri]^{2} \Big) dt^{2}\,.
\eea
 The AAI is  the total length
of the path measured using the Fubini-Study metric $S$ expressed as
$S = ({2}/{\hbar}) \int_{0}^{  t}  \Delta E (t^{\prime}) \, dt^{\prime}\,$ \cite{Anandan:1990fq}.
In the above formulas  $|\phi(t)\rangle$ is a non-stationary state with energy uncertainty $\Delta E(t) $   given by
$
\Delta E ^{2}(t) = \langle \phi(t)|H^{2}|\phi(t)\rangle -  (\langle \phi(t)|H|\phi(t)\rangle)^{2}.
$
Since the length of the  path "S" is the minimum length measured by the "reference distance" function "D" \cite{Pati1,Pati2}, the presence of the AAI in the evolution of a system implies the presence of the "D" invariant and consequentially the existence of the geometric phase (\ref{geom}).
As already mentioned in the Introduction, the AAI is produced in the evolution of systems with vacuum condensates \cite{Capolupo:2013xza} (see Appendix A), then  Eq.(\ref{geom}) suggests that vacuum condensation is also related with geometric phases.

Vacuum fluctuations may produce particle creation from vacuum or vacuum condensate in
disparate ways~\cite{Unruh:1976db}-\cite{Birrell}.
Then the geometric phase may provide the possibility to study the properties of many systems and could be used to  detect effects so far elusive to the observations, ranging from the Unruh effect to Casimir effect, including other phenomena of quantum field theory (QFT) in curved background.

As said, in the present paper we focus on the Unruh effect and on thermal states, studying the possible realization of a quantum thermometer. In our analysis we consider quantum open systems and analyze the geometric phase for mixed states in nonunitary, noncyclic evolution. In particular, we use the Wang and Liu approach \cite{Wang} and the geometric phase defined as
\begin{widetext}
\bea\label{Wang}
\Phi_{ g}  = \frac{1}{N} \sum_{k=1}^{N} \arg \lf[\sqrt{\lf[1 + b\, r(t_{0}) \varphi_{k}\ri]\lf[1 + b\, r(t) \varphi_{k}\ri]} \langle
\varphi_{k}(t_{0})|\varphi_{k}(t )\rangle \ri] - \frac{1}{N} \Im \sum_{k=1}^{N} \int_{t_{0}}^{t} \lf[1 + b\, r(t^{\prime}) \varphi_{k}\ri] \langle \varphi_{k}(t^{\prime})|\frac{\partial}{\partial t^{\prime}}|\varphi_{k}(t^{\prime}) \rangle
d t^{\prime}\,,
\eea
\end{widetext}
where the first term of the right side  represents the total phase, while the second term is the dynamic one.
In Eq.(\ref{Wang}), $r(t) $ is the Block radius, satisfying the condition $0\leq r(t)\leq 1$, $b = \sqrt{N(N-1)/2}$ and $\varphi_{k}$ are the eigenvalues of the operator $\sum^{N^{2}-1}_{i=1} n_{i}(t) \lambda_{i}$ (with $\lambda_i$ traceless and hermitian $N \times N$  operators and $n_i (t) = (N/2b) Tr \lf[\lambda_{i} \rho(t) \ri]$) which permits to write $\rho(t)$ as
$\rho(t) = (1/N) (1 +b \sum^{N^{2}-1}_{i=1} n_{i}(t) \lambda_{i})$.
Moreover,  
 $|\varphi_{k}(t ) \rangle$ and $\lambda_{k}(t) \equiv \lf[1 + b\, r(t) \varphi_{k}\ri] / N$ are   eigenstates and  eigenvalues of the density matrix $\rho$ of N-level mixed states, respectively
\bea
\rho |\varphi_{k}(t ) \rangle = \frac{\lf(1 + b\, r(t) \varphi_{k}\ri)}{N} |\varphi_{k}(t ) \rangle\,.
\eea
Then the geometric phase (\ref{Wang}) can be expressed  also as
\bea\label{Wang2}\non
\Phi_{ g} & = &  \sum_{k=1}^{N} \arg \lf[\sqrt{ \lambda_{k}(t_{0})\lambda_{k}(t))} \langle
\varphi_{k}(t_{0})|\varphi_{k}(t )\rangle \ri]
\\
& - & \Im \sum_{k=1}^{N} \int_{t_{0}}^{t}   \lambda_{k}(t^{\prime}) \langle \varphi_{k}(t^{\prime})|\frac{\partial}{\partial t^{\prime}}|\varphi_{k}(t^{\prime}) \rangle
d t^{\prime}\,.
\eea

We consider a two level open system. In this case, $N =2$, $b =1$ and the density  matrix   $\rho$ is expressed in terms of Pauli matrices $\sigma_{i}$ as
\bea\label{ro-si}
\rho(t) = \frac{1}{2}\lf(1 + \sum_{i=1}^{3} n_{i}(t) \sigma_{i} \ri)\,,
\eea
with $n_{i} = Tr (\rho \sigma_i)$, and radius of Block sphere given by $r(t) =\sqrt{n_{1}^{2}+n_{2}^{2}+n_{3}^{2} }$. Explicitly, one has
\bea\label{n1}
n_{1} &=& \rho_{12} + \rho_{21} \,,
\\\label{n2}
n_{2} &=& i (\rho_{12} - \rho_{21}) \,,
\\\label{n3}
n_{3} &=& \rho_{11} - \rho_{22}\, .
\eea
Introducing the angles such as
\bea\label{angoli}
\theta = \cos^{-1} \lf( \frac{n_{3}}{r} \ri) \,, \qquad \phi = \tan \lf( \frac{n_{2}}{n_{1}} \ri)\,,
\eea
the eigenvector $|\varphi_{1}(t )\rangle$ and $|\varphi_{2}(t )\rangle$ (apart from overall phase factors which do not contribute to $\Phi_{g}$) are
\bea
|\varphi_{1}(t )\rangle &=& \left(
                              \begin{array}{c}
                                 \cos \frac{\theta(t)}{2} \\
                                e^{i \phi(t)} \sin \frac{\theta(t)}{2} \\
                              \end{array}
                            \right)\,,
\\
|\varphi_{2}(t )\rangle &=& \left(
                              \begin{array}{c}
                                \sin \frac{\theta(t)}{2} \\
                                - e^{i \phi(t)}\cos \frac{\theta(t)}{2} \\
                              \end{array}
                            \right)\,,
\eea
and the eigenvalues of the operator $\sum_{i=1}^{3} n_{i}(t) \sigma_{i} $, are $\varphi_{1}=1$ and $\varphi_{2} = -1$.
Therefore $\lambda_{1}(t) = \frac{1}{2}[1 + r(t)]$ and $\lambda_{2} = \frac{1}{2}[1 - r(t)]$.

\section{Geometric phase and two level atoms}

We consider the interaction of an atom with vacuum modes of the electromagnetic field in the multipolar scheme \cite{Compagno} and treat the atom as an open system with a non-unitary evolution in the reservoir of the electromagnetic field.
The Hamiltonian of the  atom and the reservoir   is
\bea
H = \frac{\hbar}{2}\,\omega_{0}\,\sigma_{3}\,+\,H_{F}\,-\,  \sum_{mn}\mathbf{\mu}_{mn}\cdot \mathbf{E}(x(t))\sigma_{mn}\,,
\eea
where $\omega_{0}$ is the energy level spacing of the atom, $\sigma_{3}$ is the Pauli matrix,  $H_{F}$ is the electromagnetic field Hamiltonian, $\mathbf{\mu}_{mn}$ is the matrix element of the dipole momentum operator connecting single-particle states $u_{n}$ and $u_{n^{\prime}}$ (see \cite{Compagno}), $\sigma_{mn} = \sigma_{m }\sigma_{ n}$, and  $\mathbf{E}$ is the strength of the electric field.
Denoting by $|0\rangle$ and $\rho(0)$ the vacuum and the initial reduced density matrix of the atom, respectively,
we analyze the evolution of the total density matrix $\rho_{tot}= \rho(0)\otimes |0\rangle \langle 0|$, in the frame of the atom.
 We assume a weak interaction between atom and field, then the evolution can be written as \cite{Lind,Lind1}
 \bea\label{evolution}\non
\frac{\partial \rho(\tau)}{\partial \tau} & =& -\frac{i}{\hbar}[H_{eff},\rho(\tau)]
\\
&+&\frac{1}{2}\sum_{i,j=1}^{3} a_{i j } \lf(2 \sigma_{j} \,\rho\, \sigma_{i} - \sigma_{i}\, \sigma_{j}\, \rho - \rho \, \sigma_{i}\, \sigma_{j} \ri).
\eea
Here  $\tau$ is the proper time, $a_{i j } $ are the coefficients of the Kossakowski matrix
\bea
a_{i j } \,=\, \Sigma\, \delta_{i j } -i \Upsilon\, \epsilon_{i j k } \delta_{k 3} - \Sigma\, \delta_{i 3}\delta_{j 3},
\eea
with
\bea\label{Sigma}
\Sigma = \frac{1}{4}\lf[G(\omega_{0})+ G(-\omega_{0})\ri],\;\; \Upsilon = \frac{1}{4}\lf[G(\omega_{0})- G(-\omega_{0})\ri].
\eea
and
\bea
G(\omega)\,=\,\int_{-\infty}^{\infty} d \tau e^{i \omega \tau}G^{+}(x(\tau))
\eea
the Fourier transform of $G^{+}(x-y)$
\bea\non
G^{+}(x-y)=\frac{e^{2}}{\hbar^{2}}\sum_{i,j=1}^{3} \langle +|r_{i}|-\rangle \langle -|r_{j}|+\rangle \langle 0 |E_{i}(x) E_{j}(x)|0 \rangle\,.
\eea
$H_{eff}$ is the effective hamiltonian,
$
H_{eff}\,=\,\frac{\hbar}{2}\,\Omega\,\sigma_{3}\,,
$
where $\Omega$ is the renormalized energy level spacing containing the Lamb shift terms. Such terms can be neglected in the computation of the geometric phase, thus we   approximate the effective level spacing of the atoms $ \Omega$ with the atomic transition frequency $\omega_{0}$, i.e. $\Omega \sim \omega_{0}$.

By writing $\rho(\tau)$ in terms of Pauli matrices, as in Eq.(\ref{ro-si}),
and considering the initial state of the   atom,
 \bea
|\psi(0)\rangle = \cos \lf(\frac{\theta}{2}\ri)|+\rangle + \sin \lf(\frac{\theta}{2}\ri)|-\rangle \,,
  \eea
with $\theta \equiv \theta (0)$, one can derive the reduced density matrix $\rho(\tau)$ \cite{Capolupo:2013xza,Hu-Yu}

\begin{widetext}

\bea\label{matriceDensit}
\rho(\tau) =
\left(
  \begin{array}{cc}
    e^{-4 \Sigma \tau} \cos^{2} \lf(\frac{\theta}{2}\ri)+\frac{\Upsilon-\Sigma}{2 \Sigma}( e^{-4 \Sigma \tau} -1) & \frac{1}{2}e^{-2 \Sigma \tau -i \Omega \tau}  \sin \theta \\
  \frac{1}{2}e^{-2 \Sigma \tau + i \Omega \tau}  \sin \theta  & 1- e^{-4 \Sigma \tau} \cos^{2} \lf(\frac{\theta}{2}\ri)-\frac{\Upsilon-\Sigma}{2 \Sigma}( e^{-4 \Sigma \tau} -1)  \\
  \end{array}
\right)\,.
\eea

\end{widetext}

Denoting with $\xi(\tau)$ and $\chi(\tau)$ the following quantities
\bea\label{parametri2}
 \xi(\tau) &=& \sqrt{\chi^{2} + e^{-4 \Sigma   \tau}\sin^{2}\theta}\,,
\\\label{parametri3}
 \chi(\tau) &=& e^{-4 \Sigma \tau}\cos\theta + \frac{\Upsilon}{\Sigma}(e^{-4 \Sigma \tau}-1)\,,
 \eea
the matrix (\ref{matriceDensit}) can be written as
\bea\label{ro}
\rho(\tau) =
\frac{1}{2}\left(
  \begin{array}{cc}
    \chi + 1  &  e^{  -i \Omega \tau}  \sqrt{\xi^{2} -\chi^{2}} \\
 e^{  i \Omega \tau}  \sqrt{\xi^{2} -\chi^{2}}  &  1- \chi \\
  \end{array}
\right)\,.
\eea
Then, Eqs.(\ref{n1})-(\ref{n3}), the Block radius and the angles defined in Eqs.(\ref{angoli}) are given by
\bea
n_{1} (\tau)&=& \sqrt{\xi^{2} -\chi^{2}} \cos(\Omega \tau)\,,
\\
n_{2} (\tau) &=& \sqrt{\xi^{2} -\chi^{2}} \sin(\Omega \tau)\,,
\\
n_{3} (\tau) &=& \chi\,,
\\
r (\tau) &=& \xi\,,
\\
\theta (\tau) &=& \cos^{-1}\frac{\chi}{\xi}\,,\quad \quad \phi (\tau) = \Omega \tau\,,
\eea
respectively. The eigenvalues  of $\rho(\tau)$ are:
\bea
\lambda_{\pm} = \frac{1}{2}(1 \pm r) = \frac{1}{2}(1 \pm \xi)\,,
\eea
and the corresponding eigenvectors are
   \bea
|\phi_{+}(\tau)\rangle & = & \cos \lf(\frac{\theta(\tau)}{2}\ri)|+\rangle + \sin \lf(\frac{\theta(\tau)}{2}\ri) e^{i \Omega \tau}|-\rangle \,,
\\
|\phi_{-}(\tau)\rangle & = & \sin \lf(\frac{\theta(\tau)}{2}\ri)| +\rangle - \cos \lf(\frac{\theta(\tau)}{2}\ri) e^{i \Omega \tau}|-\rangle \,.
  \eea
We consider the initial time $t_{0} =0 $.  Being $\lambda_{-}(0) = 0$, the phase (\ref{Wang2})   becomes
\bea\label{Wang3}\non
\Phi_{ g}(t) & = &   \arg \lf[\sqrt{ \lambda_{+}(0)\lambda_{+}(t))} \langle
\phi_{+}(0)|\phi_{+}(t )\rangle \ri]
\\\non
& - & \Im   \int_{t_{0}}^{t}   \lambda_{-}(t^{\prime}) \langle \phi_{-}(t^{\prime})|\frac{\partial}{\partial t^{\prime}}|\phi_{-}(t^{\prime}) \rangle
d t^{\prime}
\\
& - & \Im   \int_{t_{0}}^{t}   \lambda_{+}(t^{\prime}) \langle \phi_{+}(t^{\prime})|\frac{\partial}{\partial t^{\prime}}|\phi_{+}(t^{\prime}) \rangle
d t^{\prime},
\eea
which reduces to
\bea\label{fase}\non
\Phi_{ g}(t) & = & \arg \lf[\cos \frac{\theta}{2} \cos \frac{\theta(t)}{2}+
\sin \frac{\theta}{2} \sin \frac{\theta(t)}{2} e^{i \Omega t} \ri]
\\
& - & \frac{\Omega}{2}\,\int_{0}^{t} \lf[ 1 - \xi(\tau) \cos \theta(\tau)\ri] d \tau\,,
\eea
with $\theta \equiv \theta (0)$.

In the next sections we consider the use of Eq.(\ref{fase}) in the detection of Unruh effect and in the building of a quantum thermometer.

\section{Unruh effect}

For an accelerated observer the ground state of an inertial system appears at a non-zero temperature depending  on the acceleration of the observer. Such a phenomenon is called the Unruh effect. It has not yet been detected. Recently it has been shown that geometric phases and invariants could allow its detection in table top
experiments~\cite{Capolupo:2013xza,Ivette1,Hu-Yu}.

Here we show that the realization of an interferometer is possible, in which paths of slightly different lengths can be chosen in order to let the geometric phase be dominating over the relative dynamical phase.
We compute the geometric phase (\ref{fase}) for the two level system in the presence of an acceleration and in the inertial case. The atom interaction with the electromagnetic field itself produces a geometric phase; however, the difference between the two phases is due only to the atom acceleration and then to the Unruh effect, since the accelerated system sees the Minkowski vacuum as a thermal Rindler vacuum.

A two-level atom uniformly accelerated in the $x$ direction with acceleration $a$, through Minkowski spacetime, is conveniently described with Rindler coordinates
$x(\tau) = \frac{c^2}{a} \cosh \frac{a \tau}{c}$, $t(\tau) = \frac{c }{a} \sinh \frac{a \tau}{c}$. For this system,
the field correlation function  is given by \cite{Capolupo:2013xza,Hu-Yu}
\bea
G^{+}(x,x') = \frac{e^{2}|\langle - |{\bf r}|+ \rangle|^{2}}{16 \pi^{2}\varepsilon_{0}\hbar c^{7}}\frac{a^{4}}{\sinh^{4}\lf[\frac{a}{2 c}(\tau - \tau'- i \varepsilon) \ri]}\,,
\eea
and its Fourier transform is
\bea
G(\omega) = \frac{\omega^{3} e^{2}|\langle - |{\bf r}|+ \rangle|^{2}}{6 \pi \varepsilon_{0}\hbar c^{3}}
\lf(1+\frac{a^{2}}{c^{2} \omega^{2}}\ri)\lf( 1+ \coth \frac{\pi c \omega}{a} \ri).
\eea
Then, the coefficients $\Sigma$ and $\Upsilon$ in Eqs.(\ref{Sigma}) become \cite{Hu-Yu,Capolupo:2013xza},
\bea
\Sigma_{a}\,=\, \frac{\gamma_{0}}{4} \,\lf(1+\frac{a^{2}}{c^{2} \omega_{0}^{2}}\ri)\,\frac{e^{2 \pi c \omega_{0}/a}+1}{ e^{2 \pi c \omega_{0}/a}-1 }
\eea
 and
 \bea  \Upsilon_{a}\,=\,\frac{\gamma_{0}}{4} \,\lf(1+\frac{a^{2}}{c^{2} \omega_{0}^{2}}\ri),
\eea
 where $\gamma_{0}\,$ is the spontaneous emission rate and $\omega_{0}$ is the atomic transition frequency.
The function  $\sin \frac{\theta(t)}{2} = \pm \sqrt{\frac{1}{2}\lf( 1 - \frac{\chi(t)}{\xi(t)} \ri)}$  in Eq.(\ref{fase}) becomes
\begin{widetext}
\bea
\sin \frac{\theta_{a}(t)}{2} & = & \pm\, \sqrt{\frac{1}{2}-\frac{R_{a}\,-\,R_{a}\,e^{4 \Sigma_{a} t} + \cos \theta}{2 \sqrt{e^{4 \Sigma_{a} t}\sin^{2} \theta\,+\,\lf(R_{a}\,-\,R_{a}\,e^{4 \Sigma_{a} t}\,+\,\cos \theta \ri)^{2}}}} \,,
\eea
\end{widetext}
and similar for $\cos \frac{\theta_{a}(t)}{2} $. Here $R_{a} = \Upsilon_{a}/\Sigma_{a}$.

For an inertial atom, $a=0$, the geometric phase $\Phi_{a=0}$ assumes the identical expression of Eq.~(\ref{fase}), with
$\sin \frac{\theta_{a}(t)}{2}$, $\cos \frac{\theta_{a}(t)}{2}$ and $\cos  \theta_{a}(t) $ replaced by
 $\sin \frac{\theta_{a=0}(t)}{2}$, $\cos \frac{\theta_{a=0}(t)}{2}$ and $\cos  \theta_{a=0}(t) $ in which the coefficients
 $\Sigma_{a}$, $\Upsilon_{a}$, $R_{a}$ are replaced by $\Sigma_{a=0}=\Upsilon_{a=0}=\gamma_{0}/{4}$, with $\gamma_0$ spontaneous emission rate, and  $R_{a=0}=1$, respectively.

The phase difference  between the accelerated and inertial atoms, $\Delta \Phi_{U}(t) =   \Phi_{a}(t) - \Phi_{a=0}(t) $, gives the geometric phase in terms of the acceleration of the atom. The value of $\Delta \Phi_{U}(t)$ depends on the ratios  $a / (c \omega_{0})$ and $\gamma_{0}/\omega_{0}$ and on the time interval $t$. Indeed  $\Delta \Phi_{U}(t)$ increases for values of the acceleration $a$ which approach to $ c \omega_{0}$, i.e. $a \sim c \omega_{0}$. Moreover,  $\Delta \Phi_{U}(t)$ is detectable when  $ \gamma_{0} / \omega_{0} > 10^{-5}$. Therefore a crucial role is played by the choice of the atomic systems used in the interferometer. Moreover, $\Delta \Phi_{U}(t)$ depends on $\theta(0)$; it reaches the maximum value for $\theta(0) = \pi /2$.
We consider an initial state with angle $\theta(0) \sim \pi /2$ and, in order to decrease the value of the acceleration, we consider the hyperfine level structure of different atoms.

In Fig.1 we plot the  difference of geometric phase $ \Delta \Phi_{U}$ as function of the acceleration $a$ for different systems.

The (blue) dashed line in the inset of the figure is obtained considering the energy splitting between the levels $F=1$ and $F=2$ of the ground state $5^{2}S_{1/2}$ of  $^{85}Rb$, (${\bf F = J + I}$ is the total atomic angular momentum, with ${\bf J}$ total electron angular momentum and ${\bf I}$ total nuclear angular momentum). For this systems one has
$\omega_{0} = 3.035 GHz$, and we considered the $D_{1}$ transition  for which $\gamma_{0}= 36.129 MHz $\cite{Daniel-Rb85}.  A similar plot is obtained if we consider the  $D_{2}$ transition for which  $\gamma_{0}= 38.117 MHz $  \cite{Daniel-Rb85}.
We derive the (red) dot dashed line in the inset by studying the energy splitting between the levels $F=1$ and $F=2$ of the $5^{2}S_{1/2}$ line of $^{87}Rb$. In this case, $\omega_{0} = 6.843  GHz$ and we consider the  $D_{1}$ transition with $\gamma_{0}= 36.129 MHz $  \cite{Daniel-Rb87}. Similar plot can be derived in the case of the $D_{2}$ transition with $\gamma_{0}= 38.117 MHz $ \cite{Daniel-Rb87}.
Such plots shows that a phase  $\Delta \Phi_{U} \sim 10^{-4} \pi$ is obtained  for  accelerations  of order of $10^{17} m/s^{2}$ and the times $t$ of order of $t \sim 1/\omega_0$, (inset of Fig.1). For such time interval, the speed of the atoms is of order of $(0.2-0.3) c$ and the spontaneous emission can be neglected, since $N(t \sim 1/\omega_0) \sim 0.99 N(0)$. The values of the phases obtained are accessible with the current technology.

Better results can be also obtained by considering other splitting between the levels of the  $^{87}Rb$, and the $6^{2}P_{1/2}$ energy splitting between the levels of the $^{133}Cs$, as shown in
 the main pictures of Fig.1.
Here, the (gray) dot dashed line represents $\Delta \Phi_{U}$ for the $5^{2}P_{1/2}$ energy splitting between the $F=1$ and $F=2$ levels of the  $^{87}Rb$. In this case,
$\omega_{0} = 814.5 MHz$ and $\gamma_{0}= 36.129 MHz $ \cite{Daniel-Rb87}.
 The (green) solid line is achieved by considering the $6^{2}P_{1/2}$ line  splitting    between  the $F=3$ and $F=4$ levels of $^{133}Cs$. For this system, $\omega_{0} = 1167.68 MHz$ and $\gamma_{0}= 28.743 MHz $ \cite{Daniel-C}.
 In these cases $\Delta \Phi_{U} \sim 10^{-4} \pi$ can be achieved for  accelerations  of order of $10^{16} m/s^{2}$ and speeds of order of $(0.2-0.3) c$, as shown in the main plots of Fig.1 and  one has $N(t \sim 1/\omega_0) \sim 0.98 N(0)$.
\begin{figure}
\begin{picture}(300,180)(0,0)
\put(10,20){\resizebox{8.0 cm}{!}{\includegraphics{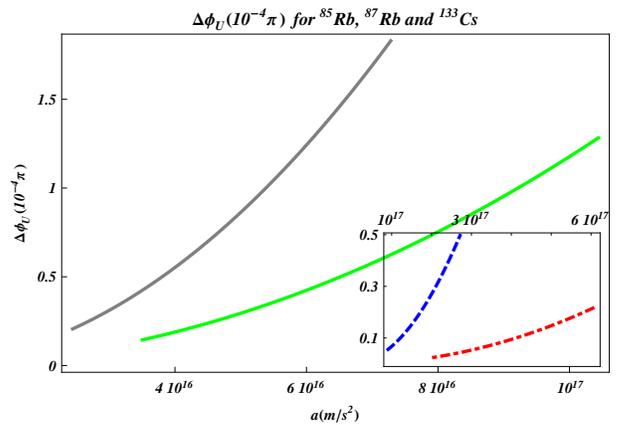}}}
\end{picture}\vspace{-1cm}
\caption{\em Plots of $\Delta \Phi_{U}$ as a function of the atom acceleration $a$, for  time intervals $t \simeq   1  /\omega_{0} $ and the splitting between the hyperfine energy levels:
main pictures: - (gray) dot dashed line: $^{87} Rb$, $5^{2}P_{1/2}$  line, splitting for $F=1 \rightarrow F=2 $ transition ($\omega_{0} = 814.5 MHz$, $\gamma_{0}= 36.129 MHz $ \cite{Daniel-Rb87}); - (green) solid line: $^{133}Cs$,   $6^{2}P_{1/2}$ line, splitting of the between  the $F=3$ and $F=4$ levels ($\omega_{0} = 1167.68 MHz$, $\gamma_{0}= 28.743 MHz $ \cite{Daniel-C}). Pictures in the inset:
- (blue) dashed line:  $^{85}Rb$, $5^{2}S_{1/2}$ line, energy splitting between the levels $F=1$ and $F=2$ ($\omega_{0} = 3.035 GHz$, $\gamma_{0}= 36.129 MHz $ for $D_{1}$ transition, $\gamma_{0}= 38.117 MHz $ for $D_{2}$ transition \cite{Daniel-Rb85}); - (red) dot dashed line:  $^{87}Rb$, $5^{2}S_{1/2}$ line, energy splitting between the levels $F=1$ and $F=2$ ($\omega_{0} = 6.843  GHz$, $\gamma_{0}= 36.129 MHz $ for $D_{1}$ transition, $\gamma_{0}= 38.117 MHz $ for $D_{2}$ transition \cite{Daniel-Rb87}).}
\label{pdf}
\end{figure}

We now analyze the characteristic of a Mach-Zehnder interferometer able to reveal the geometric phase (\ref{Wang3}) related to the Unruh effect.
The geometric phase can be detected when the dynamical phase is negligible compared with the geometric one.
The total phase, for the accelerated and for the inertial atoms, is given in terms of the geometric phase $\Phi_{g}$ by the formula
\bea\label{faseTot}
\Phi_{tot}(t) = \Phi_{g} (t) +
 \frac{\Omega}{2}\,\int_{0}^{t} [1 - \xi(\tau) \cos   \theta(\tau) ] d \tau\,,
\eea
where the second term on the right side is the dynamical phase.

We note that the dynamic phases can be made negligible compared to the geometric ones, if the branches of the interferometer are built in order that the two dynamical phases are almost equal, that is
\bea\non
\delta &=& \frac{\Omega}{2}\, \Big[ \int_{0}^{t^{\prime}} [1 - \xi_{a}(\tau) \cos   \theta_{a}(\tau) ] d \tau
\\
&-&  \int_{0}^{t} [1 - \xi_{a=0}(\tau) \cos   \theta_{a=0}(\tau) ] d \tau\Big] \ll \Delta \Phi_{U}\,.
\eea
In this case the difference of total phases $\Delta\Phi_{tot} $ detected in the cross point of the interferometer corresponds almost completely to the difference of geometric phases $\Delta\Phi$, i.e. $\Delta\Phi_{tot} \simeq \Delta\Phi$.
For example,  by considering as two level system the $5^{2}P_{1/2}$ energy splitting between the $F=1$ and $F=2$ levels of $^{87}Rb$, and an acceleration of order of $5 \times 10^{16}m/s^{2}$, one has that, in an interferometer with branches  of length of $4$ cm, the dynamical phase differences $\delta $ can be completely neglected when in a such interferometer there is a difference in the arm lengths of about  $0.1 \mu m$.

\section{ Quantum thermometer  }

In this Section we consider the interaction of an atom with thermal states. A geometric phase identical to the one in Eq.(\ref{fase}) appears also in this case. The analysis of the geometric phase (\ref{fase}) in a Mach-Zehnder interferometer could allow very precise measurement of the temperature.

For thermal states, the coefficients $\Sigma_{a}$ and $\Upsilon_{a}$ are replaced by the coefficients $\Sigma_{T}$
 and $ \Upsilon_{T}$ depending on the temperature  \cite{Capolupo:2013xza},
$\Sigma_{T}\,=\,(\gamma_{0}/{4})\,(1+{4\pi^{2} k_{B}^{2}T^{2}}/{\hbar^{2} \omega_{0}^{2}})\,({e^{E_{0}/k_{B} T}+1})/({e^{E_{0}/k_{B} T}-1})\,,$ $E_{0} = \hbar \omega_{0}$
 and
$\Upsilon_{T}\,=\,(\gamma_{0}/{4})\,(1+{4\pi^{2} k_{B}^{2}T^{2}}/{\hbar^{2} \omega_{0}^{2}})\,.$

Thus an  interferometer in which an atom follows two different paths and interacts with two thermal states at different temperatures can represent a very precise quantum thermometer.
Indeed,
if it is known the reference temperature of one thermal state, the temperature of the other one can be defined by measuring the difference between the  geometric phases generated in the two paths.

 For example, by measuring $\Delta \Phi_{T}$, one can derive precise estimations of the temperature $T_c$ of the colder source, if it is known the temperature $T_h$ of the hotter source.

 We consider the hyperfine structure of atoms and we plot in Fig.~2 $\Delta \Phi_{T}$ as function of the temperatures of cold sources  $T_{c}$, for different  $^{133} Cs$, and $^{87} Rb$ lines and  $T_{h}$ values.
 In such a figure, the (blue) dashed line is obtained for the energy splitting between the levels $F=4$ and $F=5$ of the $6^{2}P_{3/2}$ line of $^{133} Cs$. In this case $\omega_{0} = 251.09 MHz$, $\gamma_{0}= 32.889 MHz$ \cite{Daniel-C} and we considered a reference temperature $T_{h} = 10^{-2}K$.
  The (red) dot-dashed line represents  $\Delta \Phi_{T}$ for the $F=1 \rightarrow F=2 $ transition of the $5^{2}P_{1/2}$  line of $^{87} Rb$, for which     $\omega_{0} = 814.5 MHz$ and $\gamma_{0}= 36.129 MHz $ \cite{Daniel-Rb87}). A value of $T_{h} = 3 \times 10^{-2}K$ has been taken into account.
 We derive the
  (gray) solid line by considering  $T_{h} = 6 \times 10^{-2}K$ and the splitting between  $F=3 $ and $F=4$ levels of the $6^{2}P_{1/2}$  line of  $^{133} Cs$, for which $\omega_{0} = 1167.68 MHz$ and $\gamma_{0}=  28.743 MHz $ \cite{Daniel-C}.
   Moreover, we obtain the (black) dotted line by analyzing the  $F=3 \rightarrow F=4$ transition of the $6^{2}S_{1/2}$ line of  $^{133} Cs$. In this case $\omega_{0} = 9.192 GHz$, $\gamma_{0}=  28.743 MHz $ \cite{Daniel-C} and $T_{h} = 1 K$.

 The time considered are  $t \simeq  \frac{1}{4 \omega_{0}} ~s $ in order that the  particle decay can be neglected.
 The result we obtain is that, considering the hyperfine structure of the atoms, one can  measure  temperatures of the cold source of $\sim 2$ orders of magnitude below the reference temperature of the hot source.

In ref. \cite{Ivette}  the Berry phase generated by an atom coupled just to a single mode of a quantum field within a cavity is studied. Here we have studied the role of the geometric phase in the realistic case of the non unitary evolution of specific atoms interacting with thermal states. Notice that any quantum system interacting with an external field is an open system. Therefore, what it is really needed is the analysis of a geometric phase, such as the one in Eq.(\ref{fase}), which is defined  for mixed states with non-unitary evolution. Moreover, the  phase Eq.(\ref{fase}), contrarily to the Berry phase and its generalization to the mixed state \cite{Tong}, covers the case of non-cyclic and non-adiabatic evolution, therefore we are not forced to consider time intervals equal to the period, but we can consider times arbitrarily small, in order to have negligible spontaneous decay in such intervals for the energy level splitting analyzed. Also in this case, paths of slightly different lengths can be chosen in order to let the geometric phase be dominating over the relative dynamical phase.

\begin{figure}
\begin{picture}(300,180)(0,0)
\put(10,20){\resizebox{8.3cm}{!}{\includegraphics{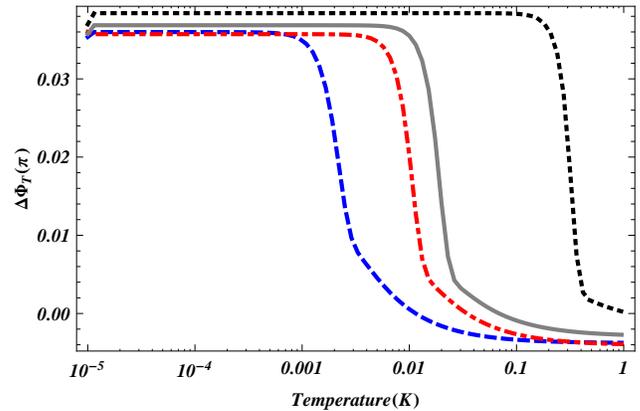}}}
\end{picture}\vspace{-1cm}
\caption{\em Plots of $\Delta \Phi_{T}$ as function of the temperatures of cold sources  $T_{c}$, for
 the splitting between the hyperfine energy levels and  $T_{h}$ values: - (blue) dashed line: $^{133} Cs$, $6^{2}P_{3/2}$ line, splitting for $F=4 \rightarrow F=5$ transition
  $(\omega_{0} = 251.09 MHz$, $\gamma_{0}= 32.889 MHz$ \cite{Daniel-C}) and $T_{h} = 10^{-2}K$; - (red)dot-dashed line: $^{87} Rb$, $5^{2}P_{1/2}$  line, splitting for $F=1 \rightarrow F=2 $ transition ($\omega_{0} = 814.5 MHz$, $\gamma_{0}= 36.129 MHz $ \cite{Daniel-Rb87}),   and $T_{h} = 3 \times 10^{-2}K$; - (gray) solid line: $^{133} Cs$, $6^{2}P_{1/2}$  line, splitting  for $F=3 \rightarrow F=4$ transition ($\omega_{0} = 1167.68 MHz$, $\gamma_{0}=  28.743 MHz $ \cite{Daniel-C}) and $T_{h} = 6 \times 10^{-2}K$; - (black) dotted line: $^{133} Cs$, $6^{2}S_{1/2}$ line, splitting  for $F=3 \rightarrow F=4$ transition ($\omega_{0} = 9.192 GHz$, $\gamma_{0}=  28.743 MHz $ \cite{Daniel-C}) and $T_{h} = 1 K$. The time considered are  $t \simeq   \frac{1}{4 \omega_{0}} ~s $. }
\label{pdf}
\end{figure}

Our results are thus realistic and new since they refer to specific atoms and their effective non-unitary evolution.

\section{Conclusions}

We have shown that all the phenomena where
vacuum condensates appear generate geometric phases in their time evolution.
In particular, we have analyzed the geometric phase for mixed state with a non-unitary evolution
 for a two level atom system.
We have shown that atoms with hyperfine structure of the energy levels, as for example $^{87}Rb$, accelerated in an interferometer with branches  of length of $4$ cm and a difference in the branch lengths of $0.1 \mu m$  could represent an efficient tool in the laboratory detection of the Unruh effect.
On the other hand, we have shown that  similar atoms, interacting with two different thermal states can be utilized in an interferometer to build  a very precise quantum thermometer.

\section*{Conflict of Interests}
The authors declare that there is no conflict of interests
regarding the publication of this paper.

\section*{Acknowledgements}

 Partial financial support from MIUR and INFN is acknowledged.

\appendix


\section{Vacuum condensate and AAI}

In this Appendix, for the reader convenience we summarize briefly how the presence of the AAI  occurs~\cite{Capolupo:2013xza} in all the phenomena in which the vacuum condensate appears \cite{Unruh:1976db}-\cite{Birrell}.
For these systems, the  physically relevant states $|\Psi (\theta)\rangle$, ($\theta \equiv \theta (\xi, t)$, with $\xi$ some physically relevant parameter) have indeed nonzero energy variance, $\Delta E(t)=\sqrt{2} \hbar\omega_{\bf k} |U_{\bf k}(\theta)| |V_{\bf k}(\theta)|$, and  AAI is given by
\bea
S(t)=2\sqrt{2}\int_{0}^{  t} \omega_{\bf k} |U_{\bf k}(\theta^{\prime})| |V_{\bf k} (\theta^{\prime})| dt^{\prime}\,,
\eea
 with $\theta^{\prime} \equiv \theta (\xi, t^{\prime})$. Here $U_{\bf k}(\theta)$ and $ V_{\bf k}(\theta)$ are the Bogoliubov  coefficients entering in the transformation $|\Psi(\theta)\rangle = J^{-1} (\theta)|\psi(t)\rangle $, with $|\psi(t)\rangle$  original state and
 $J^{-1} (\theta)$  generator of the Bogoliubov transformation,
\bea\non
 \alpha^r_{\mathbf{k}} (\theta) = J^{-1} (\theta)\,a^r_{\mathbf{k}}(t) J(\theta)
 = U_{\mathbf{k}} (\theta) \, a^r_{\mathbf{k}}(t) + V_{\mathbf{k}}(\theta) \, a^{r\dagger}_{-\mathbf{k}}( t)\,.
\eea
 $U_{\mathbf{k}} $ and $V_{\mathbf{k}}$, depend on the system one considers and satisfy the relation
$|U_{\mathbf{k}} |^2 \pm |V_{\mathbf{k}} |^2=1$, with $+$ for fermions and $-$ for bosons.

Notice that the vacuum state $|0(\theta)\rangle$ for such  systems is related to the original one $|0\rangle $ by the relation,  $|0(\theta)\rangle = J^{-1} (\theta)|0\rangle $. Therefore,
 Eq.(\ref{geom}) shows that all the phenomena characterized by the presence of modifications of vacuum fluctuations, (which are all described by Bogoliubov transformations) are also characterized by the presence of the geometric phase in their evolution.

For example, in the case of the Unruh effect, the Bogoliubov coefficients that allow to express the Minkowski vacuum in terms of Rindler states are for bosons $U_{\textbf{k}} = \sqrt{\frac{{e^{2 \pi \omega_{\textbf{k}}/a}}}{{e^{2 \pi \omega_{\textbf{k}}/a}}-1}}$  and $V_{\textbf{k}} = \sqrt{\frac{1}{{e^{2 \pi \omega_{\textbf{k}}/a}}-1}}$ and similar for fermions. Here $a$ is the acceleration of the observer. The relation between the Minkowski $|0\rangle_M$ and Rindler $|0\rangle_R $ vacua in the case of a single scalar field \cite{Crispino2008} is:
\bea
|0\rangle_M \sim \exp \lf( \frac{1}{2}\sum_{\textbf{k}} e^{-\pi \omega_{\textbf{k}}/a} a^{\dagger}_R a^{\dagger}_L\ri)|0\rangle_R ,
\eea
where $R$ and $L$ refer to modes supported in the right and left Rindler wedges respectively.

\end{document}